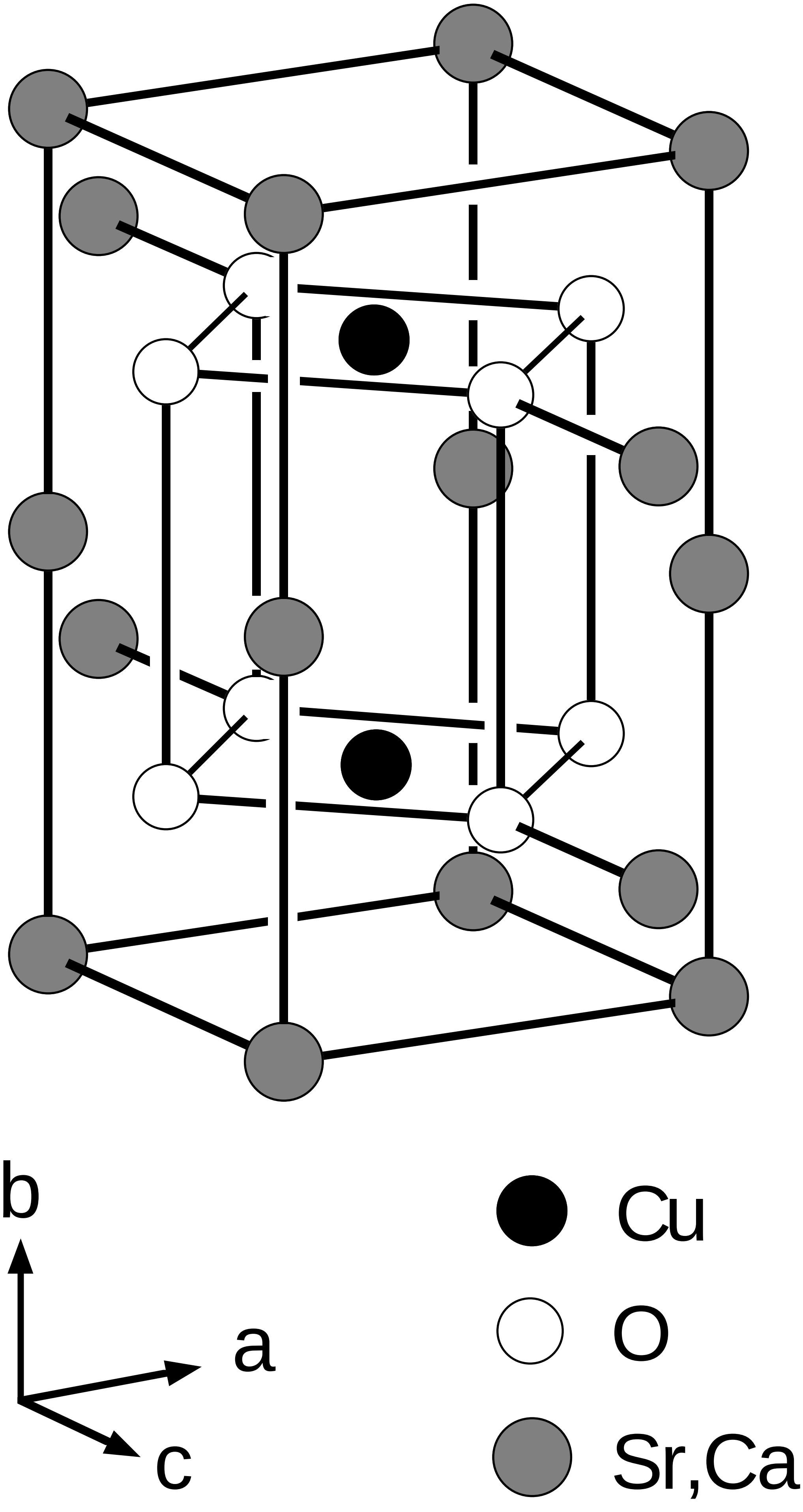

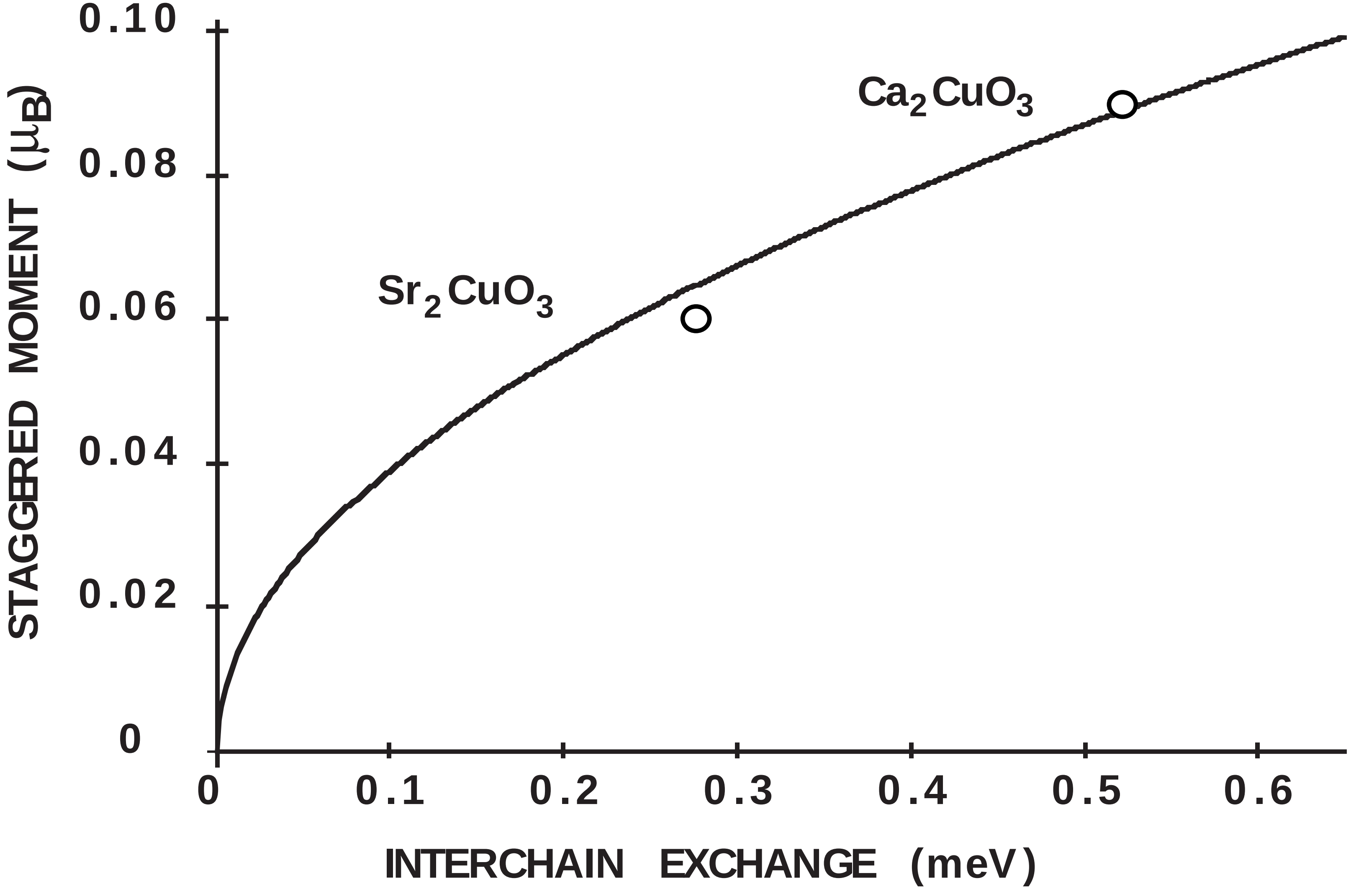

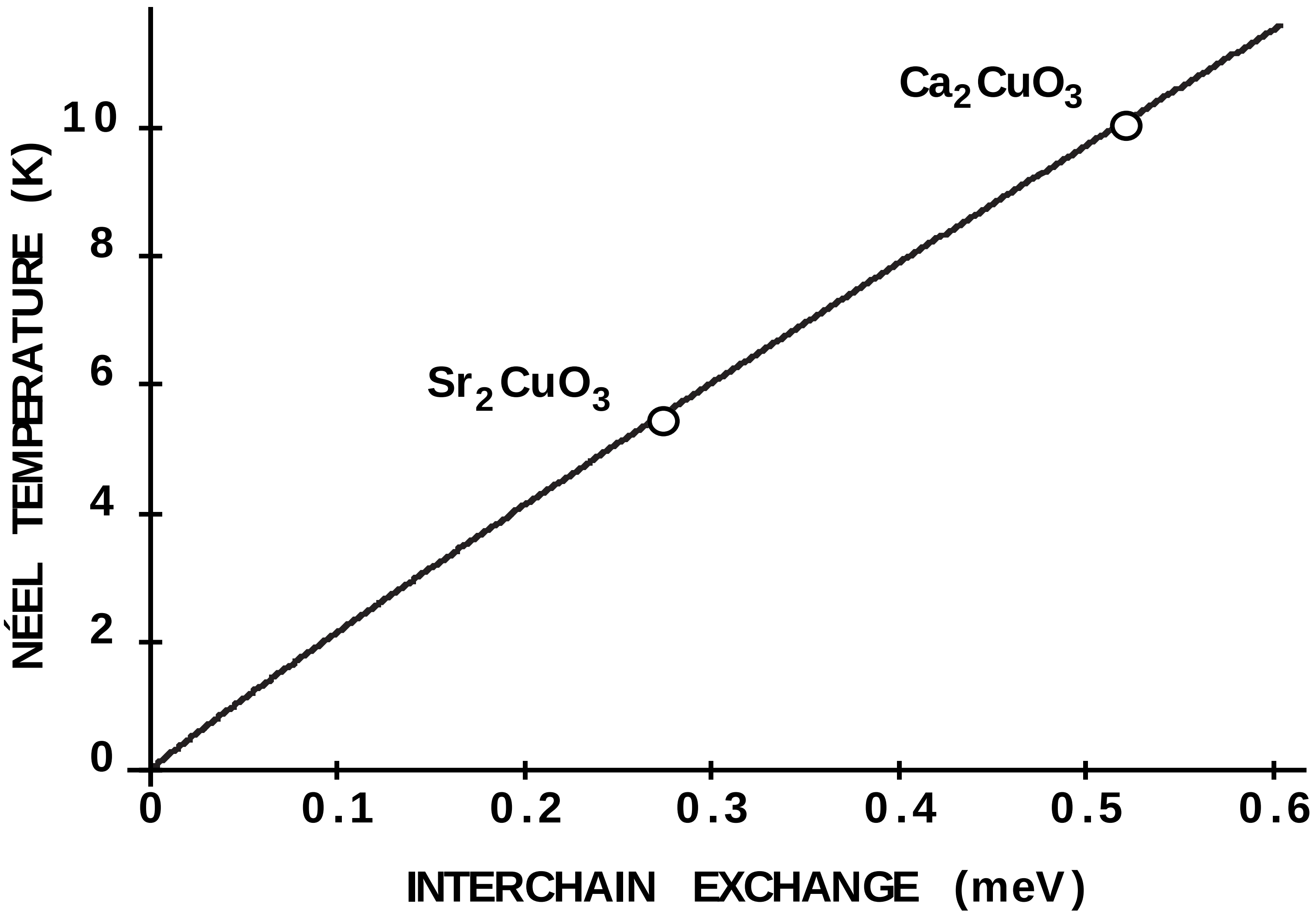

# Intra- and interchain exchange in the linear cuprate antiferromagnets $Ca_2CuO_3$ and $Sr_2CuO_3$

A. B. Van Oosten

*Max Planck Institut für Physik komplexer Systeme, Bayreutherstrasse 40, 01187 Dresden, Germany*

Ab initio quantum chemical cluster calculations of Heisenberg exchange in linear cuprate antiferromagnets are reported. For $Ca_2CuO_3$ I find $J_a$=-136 meV for the intrachain coupling and $J_b$=-0.52 meV for the interchain coupling. For $Sr_2CuO_3$ I estimate $J_a$=-136 meV and calculate $J_b$=-0.28 meV. With these values, expressions resulting from a mean field treatment of the interchain coupling reproduce the experimental values for the staggered magnetic moment and the Néel temperature $T_N$ surprisingly well. The linear dependence of $T_N$ on $J_b$ suggests that Néel order persists for any finite value of $J_b$.

PACS numbers: 71.10.-w, 71.27.+a, 74.25.Jb, 75.10.Jm, 75.30.Et

The one-dimensional (1D) spin 1/2 Heisenberg antiferromagnet is a nontrivial, but nonetheless integrable [1,2] quantum mechanical model with vanishing long-range magnetic order in the ground state. In real systems that contain antiferromagnetic chains there always exists interchain coupling. The linear cuprates $Ca_2CuO_3$ and $Sr_2CuO_3$ [3] presently attract much attention because they have exceptionally small ratio's of inter- to intrachain coupling, $J_b/J_a$, and thus are nearly ideal 1D spin 1/2 systems. This was inferred from the magnetic susceptibility and muon spin relaxation (μSR) experiments [4,5]. These materials therefore provide a unique opportunity to compare theoretical predictions with experiment, provided that the values for the coupling parameters are accurately known. Unfortunately, the values for $J_a$ as derived from experiments vary considerably. From an analysis [6,7] of magnetic susceptibility measurements [8] a value of $-J_a \approx 135\text{-}160$ meV was found. A second susceptibility study obtained a value of $J_a$=-190 meV [9]. An even larger value of J=-260 meV was estimated from the analysis of the mid-gap infrared absorption spectrum [10]. For comparison, in the planar cuprates $J_a$ varies between -100 and -140 meV [11]. For the interchain coupling $J_b$ no experimental value is available, but mean field expressions of the staggered magnetic moment [12,13] and the Néel temperature [12] in terms of $J_b$ and $J_a$ have been proposed. Recent μSR and neutron scattering experiments [5] give precise values for the staggered moment and the Néel temperature of $Ca_2CuO_3$ and $Sr_2CuO_3$. Clearly, accurate theoretical estimates for $J_a$ and $J_b$ would resolve the controversy over $J_a$ and permit a check on the validity of the mean field approximation of the interchain coupling.

In this Letter values for $J_a$ and $J_b$ are computed in a manner that is entirely independent from experiment. The computational approach that can achieve this reproduces the in-plane exchange coupling in quasi-2D cuprates within experimental accuracy [11,14]. Its application to the linear cuprates and extension to the description of interchain coupling is straightforward. The approach is an accurately balanced calculation of singlet-triplet splitting of a cluster containing two $Cu^{2+}$ ions with quantum chemical all-electron methods. Using the computed values of $J_a$ and $J_b$ a detailed comparison of the various theoretical predictions and experimental results is made.

The structure of $Ca(Sr)_2CuO_3$ consists of two interpenetrating sublattices of infinite chains of S=1/2 $[CuO_3]^{-4}$ units separated by $Ca(Sr)^{+2}$ ions [8,15], that are magnetically decoupled by symmetry. The magnetic behavior of a single sublattice can be described by the spin hamiltonian

$$H = -\Sigma_{ijk} (J_a \vec{S}_{ijk} \bullet \vec{S}_{i+1jk} + J_b \vec{S}_{ijk} \bullet \vec{S}_{ij+1k}), \quad (1)$$

where the summation runs over lattice vectors (ijk). The chains are oriented along $\vec{a}$ and the $CuO_3$-units lie in the a-c-planes, so that $J_a$ and $J_b$ are the intra- and interchain coupling, respectively.

The computational approach for the intrachain coupling $J_a$ follows that of Refs. [11,14]. As a first step the (MC)SCF singlet and triplet ground states of the planar, $D_{2h}$ symmetric, $Cu_2O_7$ cluster that occurs in the chains are calculated. It is necessary [11,14] to introduce local exchange and correlation effects on the bridging oxygen atom. This is done in a second step of the calculation through admixture of excited (MC)SCF states, that differ from the MCSCF ground states by an O -> Cu electron excitation. This is an example of non-orthogonal CI (NOCI) [16]. NOCI involves the computation of Hamilton and overlap matrix elements between determinants constructed from non-orthogonal orbital sets [17]. It has the advantage over conventional CI methods that it leads to a short, physically transparent wave function and that it is free of size consistency errors.

For the calculation of the interchain coupling $J_b$, I perform all-electron (MC)SCF calculations of the singlet and triplet ground states of $D_{2h}$ symmetric $Cu_2O_8$ and $Cu_2O_8M_{16}$ (M=Ca,Sr) clusters (Fig. 1) consisting of two $CuO_4$ units that are translated with respect to each other by one unit vector along the b-axis, so that they are located in next-nearest neighbor chains. Since there are no ligands, the exchange originates simply from the overlap between the Cu d-orbitals and $J_b$ can be found directly from the (MC)SCF ground state singlet-triplet splitting. For a study of the effect of the counterions, calculations were also performed on $Cu_2O_8M_{16}$ (M=Ca,Sr).

The clusters are embedded in a point charge environment that accurately represents the Coulomb potential inside the cluster region of a surrounding infinite lattice of formal ionic charges. The use of the simple point charges at nearest neighbor positions may lead to spurious occupation of diffuse orbitals [18], because these orbitals feel the strong attractive potential of the $Cu^{2+}$ and $Sr^{2+}$ point charges, without being repelled by the ion core [13, 16]. Therefore the potential due to the ions at the nearest neighbour positions to the cluster, is modified to be flat inside a small sphere with an ionic radius [19] around these point charges. These modified potentials improve the stability against variation of the most diffuse components of the basisset, whereas in the cluster region the potential due to the modified charges is identical to that of the bare point charges. The complete specification of the background potential is available from the authors on request.

For Cu and O the same contracted Gaussian basis sets are used as in ref. [11], namely a 14s11p6d -> 8s6p3d basis set for Cu and a 9s6p -> 3s3p one for O. It was found that $J_b$ is sensitive to extension of the Cu basis with diffuse d-functions, as will be discussed below in detail. For Ca I employed a 14s8p -> 5s2p basis set of double zeta quality [20] and a 12s6p -> 4s2p minimal basis set [21] and for Sr a (15s,9p,3d) -> (5s,3p,1d) minimal basis set [21].

First consider the intrachain coupling. Following Ref. [11], in an SCF wavefunction for the lowest triplet state of the $[Cu_2O_7]^{-10}$ model clusters two open shell orbitals occur, denoted by $d_g$ and $d_u$ and transforming as $a_g$ and $b_{3u}$, respectively. These orbitals can be considered even (g) and odd (u) linear combinations of Wannier-like orbitals, $d_1$ and $d_2$, localized at the Cu. The triplet wavefunction can be written as

$$\Psi_t = | \sigma\bar{\sigma} d_g d_u | = | \sigma\bar{\sigma} d_1 d_2 | \quad (2)$$

Here σ denotes the $O(2p_\sigma)$ orbital at the bridging oxygen, which has the same symmetry as $d_u$. For clarity all other closed shell orbitals are suppressed in the notation. The singlet corresponding to Eq.( 2) can be written as

$$\Psi_s = (2+2S^2)^{-1/2} | \sigma\bar{\sigma} ( d_1\bar{d}_2 - \bar{d}_1 d_2 ) |$$

$$= (2+2S^2)^{-1/2} \{| (1+S) \sigma\bar{\sigma} d_g\bar{d}_g - (1-S) \sigma\bar{\sigma} d_u\bar{d}_u | \}, \quad (3)$$

In Eq. (3) $d_1$ and $d_2$ have an overlap $S = \langle d_1 | d_2 \rangle$, which constitutes an additional variational parameter. As also found for the planar cuprates, the triplet and singlet SCF ground states are very well characterised by $Cu^{+2}$ $(3d^9)$ and $O^{-2}$ $(2p^6)$ and the Cu holes have almost pure $3d(x^2\text{-}y^2)$ character in a Mulliken orbital population. The singlet states (3) with optimized orbitals have $S \approx 0.04$ and their energies are 20 to 30 meV below the triplets. The wave functions (2) and (3) form an appropriate starting point of a balanced calculation of the singlet-triplet splitting [22]. At this level of the calculation the splitting corresponds to Anderson superexchange [23], which is due to charge transfer excitations of the type $d_1)^1 d_2)^1 \rightarrow d_1)^2 d_2)^0 + d_1)^0 d_2)^2$ that can only occur for the singlet.

As a next step I admix to $\Psi_t$ and $\Psi_s$ relaxed charge transfer excitations of the form $d_1)^1\sigma)^2 d_2)^1 \rightarrow d_1)^1\sigma)^1 d_2)^2 \pm d_1)^2\sigma)^1 d_2)^1$. Admixture of *unrelaxed* excitations of this type has no effect, but this is quite different for the relaxed charge transfer excitations. As discussed in full detail in Ref. [11], the charge transfer states, of $^3B_{3u}$ and $^1A_g$ symmetry, can be written as

$$\Psi_t^* = |\, d_u \bar{d}_u d_g \sigma \,| \,, \tag{4}$$

$$\Psi_s^* = \frac{(1+S^*)\,|\, d_g \bar{d}_g u_1 \bar{u}_1 \,| - (1-S^*)\,|\, d_g \bar{d}_g u_2 \bar{u}_2 \,|}{\sqrt{(2+2S^{*2})}} \,. \tag{5}$$

In the excited singlet state $\Psi_s^*$ (5) again an overlap between open shell orbitals is allowed, now between the $O(2p_\sigma)$ and the $Cu(3d_u)$ orbitals. Moreover, the open shell orbitals are allowed to mix because they have the same symmetry. It turns out that the orbitals $u_1$ and $u_2$ approximately correspond to bonding and antibonding combinations of $d_u$ and $\sigma$. It should be noted that in the wave functions (2-5) the orbitals are separately optimized. An energy separation of about 10 eV between ground and excited states is found. As for the planar cuprates, $S^* \approx 0.55$ is obtained. This result can be interpreted as a tendency towards covalent bond formation between the bridging O and the remaining $Cu^{2+}$ neighbour in the excited singlet state, which is enhanced by orbital relaxation. These effects are absent in $\Psi_t^*$. The $Cu_2O_7$ wave functions of the linear cuprates obtained here closely resemble those of the analogous clusters for the planar cuprates.

The NOCI calculation involves a 2x2 non-orthogonal diagonalization for the triplet. As argued in Ref. [11], the singlet involves a 5x5 non-orthogonal diagonalization, since S, S* and the parameter determining $u_1$ and $u_2$ as orthogonal combinations of $d_u$ and $\sigma$ have to be reoptimized. The results are displayed in Table I. For both $Ca_2CuO_3$ and $Sr_2CuO_3$ a value of $J_a$=119 meV is obtained, which is close to value for $La_2CuO_4$ reported in ref. [Van Oosten, 1996 #205]. When the 16 nearest Ca neighbors are included, as all electron atoms, the value for $Ca_2CuO_3$ increases to 136.5 meV. In view of the $Cu_2O_7$ results, it is reasonable to adopt this value also for $Sr_2CuO_3$. This value agrees with previous analyses [6,7] have given of the susceptibility data [8]. Much larger values, such as $J_a$=-190 meV [9] and $J_a$=-260 meV [10] should be excluded.

Let us now turn to the calculation of the interchain exchange $J_b$ from the singlet-triplet splitting of the $[Cu_2O_8]^{-12}$ clusters. There are no bridging O so that $J_b$ equals the singlet-triplet splitting at the (MC)SCF level and is obtained directly from wave functions analogous to equations to (2) and (3). Note that the orbital $d_u$ now transforms as $b_{1u}$ and that the σ orbital can be suppressed in the notation. Since the Cu-Cu distance of 3.278 Å ($Ca_2CuO_3$) and 3.494 Å ($Sr_2CuO_3$) is rather larger than a Cu-Cu inter-atomic distance, I investigated the effect of augmenting the basis sets with diffuse Cu d- and O p-functions. The effect of extending the d-basis is substantial and saturates after addition of $\alpha$=0.05, 0.016 and 0.005 functions. An extra $\alpha$=0.04 O p-function has negligible effect. The results are listed in Table II. The diffuse functions describe the exponential decay of the d-orbital in the region of space where the overlap is generated and contribute little to the total energy. The interchain coupling saturates towards values of $J_b$=0.33 meV ($Ca_2CuO_3$) and $J_b$=0.23 meV ($Sr_2CuO_3$). The overlaps of the d-orbitals are close to 4 ‰. The calculation was repeated for $Cu_2O_8M_{16}$ (M=Ca,Sr) clusters that include the nearest 16 counterion neighbors, using the extended Cu basis. As seen from Table III, the effect of the counterions is a 50 % increase of $-J_b$ for $Ca_2CuO_3$ and 20 % increase for $Sr_2CuO_3$, where it makes no difference wether the double zeta or the minimal basis set is used for Ca. These values can be inserted into expressions for the staggered moment and the Néel temperature obtained from a mean field treatment of $J_b$ [12,13]. It should be noted that a chain couples antiferromagnetically to two out of four next-nearest neighbors, whereas the coupling to the other two ($J_c$) is ferromagnetic and should be considerably smaller than $J_b$. The coupling to the four nearest neighbors cancels in the ordered state. As can be seen from Table IV, this yields values in excellent agreement with experiment [5]. Equally good agreement exists between the measured values and the mean field prediction for $T_N$ [12], when the calculated $J_a$ and $J_b$ are used. The situation is graphically displayed in Fig. 2. In view of the small number of neighbor chains, it is quite surprising that mean field theory should perform so well in the present case. The linear dependence $T_N$ as a function of $J_b$ at $J_b/J$ as small as $2.4 \times 10^{-3}$ supports the idea that long range magnetic order persists at arbitrarily small coupling ratio [24].

In summary, the intra- and interchain magnetic coupling, $J_a$ and $J_b$, in $Ca_2CuO_3$ and $Sr_2CuO_3$ were computed . I obtain values of $J_a$=-119 meV for the intrachain coupling in both $Ca_2CuO_3$ and $Sr_2CuO_3$, respectively, with $Cu_2O_7$ clusters. This value is close to the value previously obtained for $La_2CuO_4$. For the $Cu_2O_7M_{16}$ (M=Ca,Sr) clusters $J_a$ is enhanced to -136 meV, which agrees with several analyses of the magnetic susceptibility of $Sr_2CuO_3$. For the interchain coupling I obtain $J_b$=-0.52 meV and $J_b$=-0.28 meV, respectively. With these values mean field expressions predicts magnetic moments and Néel temperatures in close agreement with experimental data. This constitutes strong support to the mean field treatment of the interchain coupling and to the idea that magnetic order occurs for any finite interchain coupling.

I am grateful to Stephan Drechsler, Walter Stephan and Karlo Penc for stimulating discussions.

# Tables

| | a | b | c |
|---|---|---|---|
| $Sr_2CuO_3$ ($Cu_2O_7$) | -32.8 | -118.7 | -100..-260[1] |
| $Ca_2CuO_3$ ($Cu_2O_7$) | -36.5 | -118.8 | —— |
| $Ca_2CuO_3$ ($Cu_2O_7Ca_{16}$) | -48.1 | -136.5 | —— |

Table I. Calculated intrachain coupling $J_a$ of the linear cuprates $Sr_2CuO_3$ and $Ca_2CuO_3$: a) Triplet-singlet splitting of Eqs. (2,3); b) Non-orthogonal CI involving (2,3) and (4,5); d) Experiment. [1]Refs. [6-10].

| added diffuse exponent | | $J_b$ (meV) | |
|---|---|---|---|
| O 2p | Cu 3d | $Sr_2CuO_3$ | $Ca_2CuO_3$ |
| —— | —— | –0.167 | –0.209 |
| —— | 0.05 | –0.176 | –0.252 |
| —— | 0.05, 0.016 | –0.226 | –0.321 |
| —— | 0.05, 0.016, 0.005 | –0.228 | –0.330 |
| 0.04 | —— | –0.166 | –0.213 |

Table II. Calculated interchain coupling $J_b$ of $Sr_2CuO_3$ and $Ca_2CuO_3$ from $Cu_2O_8$ clusters, for various basis sets.

| cluster | $J_b$ (meV) | |
|---|---|---|
| O 2p | $Cu_2O_8$ | $Cu_2O_8M_{16}$ |
| $Sr_2CuO_3$ | –0.228 | –0.276 |
| $Ca_2CuO_3$ | –0.330 | –0.522 |

Table III. Calculated interchain coupling $J_b$ of $Sr_2CuO_3$ and $Ca_2CuO_3$ for $Cu_2O_8$ and $Cu_2O_8M_{16}$ (M=Sr,Ca) clusters.

| | magnetic moment ($\mu_B$) | | Néel temperature (K) | |
|---|---|---|---|---|
| | Mean field[1,2] | μSR[3] | Mean field[2] | μSR n-diffraction[3] |
| $Sr_2CuO_3$ | 0.065 | 0.06(1) | 5.58 | 5.41(1) |
| $Ca_2CuO_3$ | 0.089 | 0.09(1) | 10.12 | 10 |

Table IV. Mean field and experimental staggered magnetic moment and Néel temperature in $Sr_2CuO_3$ and $Ca_2CuO_3$, using $Cu_2O_8M_{16}$ (M=Sr,Ca) clusters. [1]Ref. [13]. [2]Ref. [12]. [3]Ref. [5].

## Figure captions

FIG. 1. The $Cu_2O_8M_{16}$ (M=Sr,Ca) cluster.

FIG. 2. Staggered magnetic moment versus interchain coupling $|J_b|$. The solid curve represents the mean field expression. The points correspond to Table IV.

FIG. 3. Néel temperature versus the interchain coupling $|J_b|$. The solid curve represents the mean field expression. The points correspond to Table IV.